\documentclass[aps,prb,twocolumn,showpacs,floatfix]{revtex4}
\usepackage{graphicx}
\usepackage{dcolumn}
\usepackage{amsmath,amssymb,amsthm}

\usepackage{color}

\begin{document}

\title{Dynamics and relaxation of $sp$ biexcitons in disk-shaped quantum dots}
\author{V. Moldoveanu$^{1}$, I. V. Dinu$^1$ and R. Dragomir$^{1,2}$}
\address{$^1$ National Institute of Materials Physics, PO Box MG-7, Bucharest-Magurele,Romania}
\address{$^2$ Faculty of Physics, University of Bucharest, Romania}
\date{\today}

\begin{abstract}

We study the effects of intraband relaxation processes on optical manipulation protocols for $sp$ biexcitons
hosted by CdTe disk-shaped quantum dots. The many-body states are calculated within the configuration
interaction method starting from single-particle states given by the $k\cdot p$ theory.
The time-dependent occupations of relevant many-body states are extracted from the von Neumann-Lindblad equation for the density operator.
We mainly investigate the generation of $sp$ biexcitons with two pulses of different polarizations $\sigma_+$ and $\sigma_-$.
The fast hole relaxation processes prevent a high-fidelity controlled operation on $sp$ biexcitons and lead to the
occupation of some transient states which can be optically probed. More importantly, the many-body structure of the transient states
consists of two holes on the $s$ shell and antiparallel $sp$ triplet states for electrons. Our simulations show that these
triplet states are more stable against decoherence as they can only be damaged through slow electron relaxation.
The configuration mixing due to correlation effects is also discussed.

\end{abstract}


\maketitle

\section{Introduction}

The on-demand generation and the ultrafast control of electron-hole pairs in self-assembled
quantum dots (QDs) represent ultimate steps towards excitonic quantum hardware \cite{Ramsay1,Rossi}.
  Two-qubit operations based on exciton-biexciton transitions were observed \cite{Stufler,Li}
and conditional operations in single or double quantum dots theoretically studied more than a decade ago
\cite{Biolatti1,Biolatti2,Troiani} are now demonstrated experimentally \cite{Boyle,Gamouras,Mathew,Zecherle}
by using femtosecond pulse shaping techniques. The implementation of one or two exciton quantum gates essentially relies on
 the control of many-body states (MBS) via optical Rabi oscillations \cite{Stievater,Ramsay2}.
Then the simplest theoretical approach to exciton dynamics builds on two-level models
borrowed from quantum optics in which the many-body structure and the geometry are usually disregarded.

Nonetheless, the Coulomb interaction, the confinement potential
and carrier relaxation unavoidably play a major role in self-assembled QDs.
The existing literature provides extensive and realistic calculations of exciton and biexciton states and absorption spectra
for various QD shapes (see e.g Ref.\onlinecite{Narvaez}) but systematic {\it time-dependent} calculations derived from the Fock space structure
of the exciton complexes are still required. Recent studies emphasized the effect of Coulomb correlations on carrier
scattering and therefore the need to go beyond the single-particle framework of the Boltzmann equation \cite{Steinhoff}.

On the other hand the efficiency of any quantum protocol for optical manipulation of QD many-body states is
strongly affected by intraband relaxation (IBR). Therefore predictive theoretical simulations must be performed
by taking such processes into account. As expected, the intraband relaxation can be safely neglected
if one restricts the optical driving to $s$-shell excitons and biexcitons which are rather robust
against decoherence. In contrast, the intraband relaxation is crucial for an appropriate description of the $p$-shell dynamics.
Actually various effects due to the simultaneous optical driving of $s$ and $p$ shells were already reported in recent experiments which
partly motivated the present work.

Transitions between $p$ shells in InGaAs QDs have been measured by Seidl {\it et al.} \cite{Seidl} to capture interaction effects,
as the $p$ shell is filled by four electrons. Kazimierczuk {\it et al.} \cite{Kazimierczuk1,Kazimierczuk2} investigated
the recombination of doubly negatively charged excitons in CdTe/ZnTe QDs and estimated the electron-electron exchange interaction
from the emission lines of the $sp$ singlet and triplet states.
The  non-trivial effect of intraband relaxation processes and spin Coulomb blockade on optical properties of $sp$ trions
has been investigated both experimentally \cite{Sotier} and theoretically \cite{Kavousanaki,Huneke}.
The control of spin orientation by intersubband pulses in quantum wells was also proposed in the
work of Vogel {\it et al.} \cite{Vogel} using the momentum space single-particle density-matrix approach.

The aim of this work is to address theoretically the dynamics of $sp$ biexcitons in disk-shaped quantum dots in the presence
of intraband relaxation.
The valence-hole states are obtained from the multiband Luttinger-Kohn $k\cdot p$
theory \cite{Lok} and therefore we can also deal with QDs in which light-holes (LH) and heavy-holes (HH) are mixed \cite{Climente}.
We obtain the exciton and biexciton states in the presence of both interband and intraband Coulomb interaction.
Our main focus is on the so-called controlled operations on $sp$ biexcitons which were not studied so far.
In fact previous experimental and theoretical work payed more attention to manipulation schemes for $s$-type biexcitons
(see e.g Refs. [\onlinecite{Boyle,Mathew,Chen}] and references therein).

The content of the paper goes as follows: the model and the theoretical tools are presented in Section II,
the numerical results are discussed in Section III, Section IV being left to conclusions.

\section{Formalism}
In order to discuss the optical driving of $s$ and $p$ shells one should go beyond the simple two-band model for electrons and holes.
We therefore describe the holes confined in disk-shaped quantum dots
by the $4\times 4$ Kohn-Luttinger (KL) Hamiltonian, in which the wavevector ${\bf k}$ is replaced by
the momentum operator $-i\hbar {\bf \nabla}$ as requested by the multiband $k\cdot p$ theory \cite{Lok}.
The resulting operator-valued matrix reads as follows (see e.g. Ref.\onlinecite{Climente1}):
\begin{equation}
\hat{h}_{KL}=
 \begin{pmatrix}
 \hat{P}_{+} & \hat{R} & -\hat{S} & 0\\
 \hat{R}^{*} & \hat{P}_{-} & 0 & \hat{S}\\
 -\hat{S}^{*} & 0 & \hat{P}_{-} & \hat{R}\\
 0 & \hat{S}^{*} & \hat{R}^{*} & \hat{P}_{+},
\end{pmatrix},
 \label{LuttingerKohn}
\end{equation}
where we introduce the notations:
\begin{eqnarray}
 \hat{P}_{\pm}&=&\frac{\hbar^{2}}{2 m_{0}}\left ( \left(\gamma_{1}\pm\gamma_{2}\right)\hat{p}_{\bot}^{2}+\left(\gamma_{1}\mp2 \gamma_{2}\right)\hat{p}_{z}^{2} \right )\\
\hat{R}&=&-\sqrt{3}\frac{\hbar^{2}}{2m_{0}}\gamma_{2}\hat{p}_{-}^{2},\quad \hat{S}=2\sqrt{3}\frac{\hbar^{2}}{2 m_{0}}\gamma_{3}\hat{p}_{-} \hat{p}_{z}.
\end{eqnarray}
$\gamma_{i}$ are the Luttinger parameters and $m_0$ denotes the free electron mass.
Here $\hat{p}_{z}=-i \nabla_{z}$, $\hat{p}_{\pm}=-i\left(\nabla_{x}\pm i \nabla_{y}\right)$ and $\hat{p}_{\bot}^{2}=\hat{p}_{x}^{2}+\hat{p}_{y}^{2}$.
For simplicity we consider vanishing Dirichlet boundary conditions on the disk of radius $R$ and
height $W$.
The valence band wavefunctions are found by diagonalizing the KL Hamiltonian w.r.t single-particle basis
$\{|J_z,m_z,n,l\rangle \}$, where $|J_z,m_z,n,l\rangle:=|J_z\rangle | \phi_{m_znl} \rangle$, $|J_z \rangle $
being the band-edge Bloch functions ($J_z=\pm 3/2,\pm 1/2$) and $\phi_{m_znl}$ the appropriate envelope functions for
cylindrical confinement ($\rho\in [0,R]$, $z\in [-W/2,W/2]$):
\begin{equation}
\phi_{m_znl}(\rho,\theta,z)=\frac{e^{im_z\theta}}{\sqrt \pi R}\cdot\frac{J_{m_z}(\alpha_n^{m_z}\rho/R)}
{|J_{m_z+1}(\alpha_n^{m_z})|}\cdot\xi_l(z).\label{Bessel}
\end{equation}
Here $\alpha_n^{m_z}$ is the $n$-th zero of the Bessel function $J_{m_z}$ of order $m_z$ and
$\xi_l$ are eigenfunctions associated to the vertical confinement, i.e. $\xi_l(z)=\sqrt{\frac{2}{W}}\sin\left(\frac{\pi lz}{W}\right )$
for $l$ even and $\xi_l(z)=\sqrt{\frac{2}{W}}\cos\left(\frac{\pi lz}{W}\right )$ for $l$ odd.
 The eigenstates of $\hat{h}_{KL}$ are written as linear combinations of basis vectors $|J_z,m_z,n,l\rangle$:
\begin{equation}\label{spinors}
|\psi^v_{i,F_z}\rangle = \sum_{J_z+m_z=F_z}\sum_{n,l}C^{i,F_z}_{n,l}|\phi_{m_znl}\rangle |J_z \rangle.
\end{equation}
Note that for any $i$ the total orbital quantum number $F_z=J_z+m_z$ is conserved and that there could be more eigenfunctions
with the same $F_z$. The energies in the valence band (VB) are denoted by $E^v_j$ and the hole energies are defined as $E_j^h=-E_j^v$.
The index $j$ orders the levels such that $E_1^h\le E_2^h\le E_3^h...$ and the quantum number $F_z$ associated to a given $j$ is not
indicated unless necessary. If the light-hole heavy-hole mixing is small a given Luttinger spinor can be specified by a single pair of quantum
numbers $\{J_z, m_z\}$. In this case we shall simply denote the corresponding state as
$|J_z,m_z\rangle$  (see Section III).

The conduction band (CB) electrons are described by a single-band effective mass Hamiltonian $\hat{h}_C$
which is also diagonalized in the basis $\{|S_z=\pm 1/2,m_z,n,l\rangle\}$ in order to obtain the single-particle
states and energies $|\psi^c_{i,S_z}\rangle$ and $E^c_i$.


Then one can write down the fully interacting Hamiltonian in terms of creation/annihilation operators for electrons
($a_{i}^{\dagger},a_{i}$ ) and holes ($b_{j}^{\dagger},b_{j}$):
\begin{eqnarray}\nonumber
H_0&=&\sum_{i=1}^{N_C}E_{i}^{c}a_{i}^{\dagger}a_{i}+\sum_{j=1}^{N_V}E_{j}^{h}b_{j}^{\dagger}b_{j}
+\frac{1}{2}\sum_{i,j,k,l} a_{i}^{\dagger}a_{j}^{\dagger}a_{l}a_{k}V_{ijkl}^{ee}\\\nonumber
&+&\frac{1}{2}\sum_{i,j,k,l} b_{i}^{\dagger}b_{j}^{\dagger}b_{l}b_{k}V_{lkji}^{hh}
-\sum_{i,j,k,l}a_{i}^{\dagger}b_{l}^{\dagger}b_{j}a_{k}V_{ijkl}^{eh}\\\label{HHzero}
&=&H_{KL}+H_C+W^{ee}+W^{hh}+W^{eh},
\end{eqnarray}
where $H_{KL}$ and $H_C$ are 2nd quantized versions of single particle Hamiltonians ${\hat h}_{KL}$ and ${\hat h}_C$ and
 we introduced the electron-electron ($W^{ee}$), electron-hole ($W^{eh}$) and hole-hole
($W^{hh}$) interactions. For computational reasons the single-particle indices $i$ and $j$ were restricted to
$N_C$ and $N_V$. Otherwise stated, we shall construct a relevant set of MBS starting from the $N_C$ ($N_V$) single-particle states
with the lowest energy for electrons (holes). We checked that the numerical results do not change significantly when $N_C$ and $N_V$ increase.
The Coulomb potential is given as usual. For example:
\begin{equation}
V_{ijkl}^{ee}=\int d{\bf r}\int d{\bf r}'\psi^*_i({\bf r})\psi_k({\bf r})
\frac{e^2}{4\pi\epsilon_0\epsilon_r|{\bf r}-{\bf r'}|}\psi^*_j({\bf r'})\psi_l({\bf r'}),
\end{equation}
where $\epsilon_0$ and $\epsilon_r$ are the vacuum and relative permittivities. We calculate these terms
by appropriate numerical methods, taking as well into account the Coulomb selection rules \cite{Rego}.
Note that the Coulomb matrix elements $W^{eh}$ and $W^{hh}$ contain complicated Luttinger spinors (see Eq.(\ref{spinors})).

The interacting states are calculated numerically by diagonalizing $H_0$ in the Fock space of the noninteracting
MBS $\{\nu\}_{\nu=1,N_{MB}}$ which solve $(H_{KL}+H_C)|\nu\rangle={\cal E}^{(0)}_{\nu}|\nu\rangle $
($N_{MB}$  the number of non-interacting many body configuration kept in the diagonalization procedure):
\begin{equation}\label{Hzero}
H_0=\sum_{\nu}{\cal E}^{(0)}_{\nu}|\nu\rangle\langle\nu |+\sum_{\nu,\nu '}W_{\nu\nu'}|\nu\rangle\langle\nu'|.
\end{equation}
$W_{\nu\nu'}$ are the matrix elements of the total interaction term $W=W^{ee}+W^{hh}+W^{eh}$.
The set of non-interacting MBS is truncated according to the optical processes
one is interested in. Typically this set contains the ground state (i.e. the VB is completely filled
while the CB is empty) and all exciton and biexciton states. Note that this selection of the relevant
states implies that $N_{MB}<2^{N_C+N_V}$.
The interacting states are denoted by bold letters and are obtained as linear combinations of non-interacting configurations:
\begin{equation}\label{IMBS}
 |{\boldsymbol\nu }\rangle=\sum_{\nu'}C^{{\boldsymbol\nu}}_{\nu'}|\nu'\rangle.
\end{equation}
Finally, the light-matter Hamiltonian reads as:
\begin{equation}
V_R(t)=\frac{eA_0(t)}{m_0}\sum_{i\in C,j\in V}\left ((e^{-i\omega t}S^+_{ij}+e^{i\omega t}S^-_{ij})a^{\dagger}_ib_j^{\dagger}+h.c \right ),
\end{equation}
where $A_0(t)$ is the pulse envelope and $\omega$ is its frequency. For simplicity we consider monochromatic pulses and
real rectangular envelopes $A_0$ but one could deal with more complex shapes as well.
The electric field associated to the vector potential $A_0$
is denoted by $F$.
 $S^{\pm}_{ij}$ are the interband optical coupling matrix
elements ($\vec{e}_{\sigma_{\pm}}$ are right/left circular polarization vectors):
\begin{equation}\label{SS}
S^{\pm}_{ij}=\langle \psi_i^c,(\vec{p}\vec{e}_{\sigma_{\pm}})\psi_j^h \rangle, \quad i\in CB, j\in VB.
\end{equation}
Note that $S^+_{ij}=(S^-_{ji})^*$. Replacing the conduction and valence band single-particle states in
Eq.(\ref{SS}) one can calculate $S^{\pm}_{ij}$ in terms of the parameter $P=\frac{i\hbar}{m_0}\langle s,p_x X\rangle$ which in turn
is related to the Kane energy $E_p=\frac{2}{m_0}|\langle s,p_x X\rangle|^2=\frac{2m_0}{\hbar^2}P^2$ (see e.g. Ref.\onlinecite{Lok}).

The dynamics of the system is derived from the density operator $\rho$
which obeys the von Neumann-Lindblad equation (see e.g Ref.(\onlinecite{Kavousanaki})):
\begin{equation}
i\hbar {\dot\rho}(t)=[H_0+V_R(t),\rho(t)]+\sum_{\lambda }{\cal L}_{\lambda}[\rho(t)].
\end{equation}
Here ${\cal L}_{\lambda}$ describes the intraband relaxation in the band $\lambda$
($\{,\}$ denotes the anticommutator):
\begin{eqnarray}
{\cal L}_{\lambda}[\rho(t)]=\sum_{i,j\in \lambda}\gamma_{ij}
\left\lbrace X_{ij}^{\lambda}\rho(t) X_{ij}^{\lambda\dagger}-\frac{1}{2}
\{X_{ij}^{\lambda\dagger}X_{ij}^{\lambda},\rho(t)\}\right\rbrace.
\end{eqnarray}
The relaxation rates $\gamma_{ij}$ are associated to a pair of single-particle states in the band $\lambda$.
The jump operators $X_{ij}^{\lambda}$ are defined as:
\begin{equation}
X_{ij}^{c}=a^{\dagger}_ia_j,\quad X_{ij}^{v}=b^{\dagger}_ib_j.
\end{equation}
In this approach the intraband relaxation rates are input (phenomenological) parameters.
The values considered in the numerical simulations are within the range of measured relaxation rates.
The spin-conserving hole relaxation is the fastest process described by the relaxation time for holes
$\tau_h=\gamma_{ij}^{-1}$ where $i,j$ are hole levels having the same spin. $\tau_h$
is about few picoseconds or even less than 1 ps. Spin-conserving relaxation time for electrons
$\tau_e$ usually exceeds 10 ps and can go up to 100ps (see e.g. Ref.\,\onlinecite{Vogel}) if the gap between the $s$ and $p$ shells is less than the energy of the
longitudinal optical phonons. The spin-flip processes in the conduction band are much slower so we take the corresponding
relaxation time $\tau_s=10\tau_e$. The hole spin relaxation is neglected as the measured values \cite{Flissikowski} are of order of nanoseconds.

The Lindblad equation is solved numerically on the Fock subspace of the interacting MBS.
The diagonal elements of the reduced density operator provide the population of a given many-body configuration. Therefore
we introduce the notation $P({\boldsymbol\nu})=\langle{\boldsymbol\nu}|\rho(t) |{\boldsymbol\nu}\rangle$.

\section{Numerical results}
\subsection{Many-body configurations in disk-shaped QDs}

Let us start with a brief spectral characterization of disk-shaped QDs.
The quantum dots considered here have a rather small aspect ratio $W/2R$.
In this case the LH-HH mixing is small and the Luttinger spinors associated to the highest levels in the valence
band are mostly heavy-hole like (see e.g Ref.\onlinecite{Climente1}). The single-particle spectra for electrons and
holes are obtained by diagonalizing the corresponding Hamiltonians, as discussed in Section II. It turns out that
the low-energy hole spectrum is accurately computed if the single-particle basis $\{|J_z,m_z,n,l\rangle \}$ is
truncated to $m_z=-3,..,3$, $n_{{\rm max}}=7$ and $l_{{\rm max}}=8$.
The first ten hole levels are shown in Fig.\,1(a) for CdTe quantum dots of different radii $R$.
The lowest energy Kramers doublet is the so-called $s$ shell, the associated states being denoted by
$|J_z=\pm\frac{3}{2},m_z=0\rangle$. The next two doublets belong to $p$ shell (i.e $m_z=\pm 1$) and are also mostly
HH states so we denote their eigenstates by $|J_z=\pm\frac{3}{2},m_z=\pm 1\rangle$.
For CdTe QDs considered here the spin split-off states are well separated from LH/HH states \cite{Tyrell} such that
 their contribution can be safely neglected.

Fig.\,1(a) also shows that the $s$ and $p$ shells are separated by a gap $\Delta_{sp}$ while a much smaller gap $\Delta_{pp}$ exists between the
two doublets in the $p$ shell. In  Fig.\,1(b) we give the dependence of these gaps on the radius $R$ for fixed height $W=5$nm.
As $\Delta_{sp}$ is rather large one can argue that an optical pulse centered on the $s$ shell
will not drive electrons from the $p$ shells. On the contrary, the small values for $\Delta_{pp}$
allow simultaneous activation of two states from the $p$ shell provided they obey the same selection rules.
These results show that the $sp$ biexciton states should be built from the six single-particle states belonging to the
$s$ and $p$ shells. This truncation is expected to give a good description of the dynamics as long as the $d$ shell does not
become optically active when a laser pulse creates holes in the $p$-shells.

 The finite splitting $\Delta_{pp}$ within the $p$-shell is due to light-hole heavy-hole mixing induced by the spin-orbit coupling.
In this case the Lutinger spinors (see Eq.(\ref{spinors})) have a dominant heavy-hole component corresponding to $m_z=1$ or $m_z=-1$
but also some minor components with different $m_z$. The splitting between the $p$-shell subbands is due to the different centrifugal
energies of these minor components (for more details see Ref.(\onlinecite{Climente}).

\begin{figure}[tbhp!]
\hskip 1cm
\includegraphics[angle=0,width=0.6\textwidth]{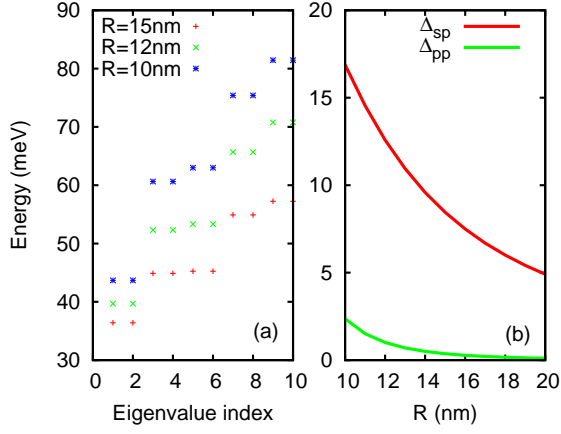}
\caption{(Color online) (a) The lowest ten hole levels of CdTe QDs of different radii. The energies are
measured w.r.t $E_g/2$ (the bulk gap for CdTe $E_g=1.49$eV). (b) The gap $\Delta_{sp}$ and the gap $\Delta_{pp}$ within the $p$ shell as a function of
radius $R$. The QD height $W=5$nm and the Luttinger parameters for CdTe
are $\gamma_1=5.37$, $\gamma_2=1.67$, $\gamma_3=1.98$.}
\label{fig1}
\end{figure}

The $p$-shell states in the conduction band are fourfold degenerate.
We introduce the following notations for the single-particle electron states:
$|\uparrow_{s}\rangle =|S_z=\frac{1}{2},m_z=0\rangle$, $|\downarrow_{s}\rangle =|S_z=-\frac{1}{2},m_z=0\rangle$,
$|\uparrow_{p_{\pm}}\rangle =|S_z=\frac{1}{2},m_z=\pm 1\rangle$ and $|\downarrow_{p_{\pm}}\rangle =|S_z=-\frac{1}{2},m_z=\pm 1\rangle$.
Note that the orthogonality of the envelope functions leads to additional optical selection rules which
do not allow the coupling between the $s$ and $p$ shells.

We describe the non-interacting MBS built from single-particle states by the quantum
numbers of the electrons occupying the conduction band and by the quantum numbers of the
holes. Recalling that $J_z^{h}=-J_z$ and $m_z^{h}=-m_z$ one can set
notations for the hole $s$-shell states $|\Uparrow_s\rangle=|J_z^{h}=\frac{3}{2},m_z^h=0\rangle $,
$|\Downarrow_s\rangle=|J_z^{h}=-\frac{3}{2},m_z^h=0\rangle $ and for the $p$-shell states
$|\Uparrow_{p_{\pm}}\rangle:=|J_z^{h}=\frac{3}{2},m_z^h=\pm 1\rangle$
and $|\Downarrow_{p_{\pm}}\rangle:=|J_z^{h}=-\frac{3}{2},m_z^h=\pm 1\rangle$.
The many-body states are then described by combining the above
notations for electrons and holes. For example, the non-interacting $s$-shell electron-hole pairs read as
$|\uparrow_s\Downarrow_s\rangle$ and $|\downarrow_s\Uparrow_s\rangle$ whereas the $p$-shell states are defined
as $|\downarrow_{p_{\pm}}\Uparrow_{p_{\mp}}\rangle $ and $|\uparrow_{p_{\pm}}\Downarrow_{p_{\mp}}\rangle $.

Before showing results on exciton dynamics we find instructive to discuss the effect of the electron-hole interaction
on the so-called configuration mixing due to the off-diagonal matrix elements of the Coulomb interaction
$\langle\nu|W^{eh}|\nu^{\prime}\rangle$ (see e.g. Ref.\onlinecite{Biolatti2}).
For simplicity it is oftenly argued that as long as the Coulomb interaction is much smaller
than the confinement energies the mixing of various non-interacting MBS $\{|\nu\rangle\}$ (see Eq.(\ref{IMBS})) can
be safely neglected. At the formal level this approximation implies that the contribution of the interaction Hamiltonian
comes only from terms which conserve the occupation number of each single-particle level \cite{Kavousanaki}.
On the contrary, if the non-conserving terms are comparable or even larger
than the level spacing one cannot disregard their contribution and the configuration mixing has to be taken
into account.

\begin{figure}[tbhp!]
\hskip -2cm
\includegraphics[angle=0,width=0.45\textwidth]{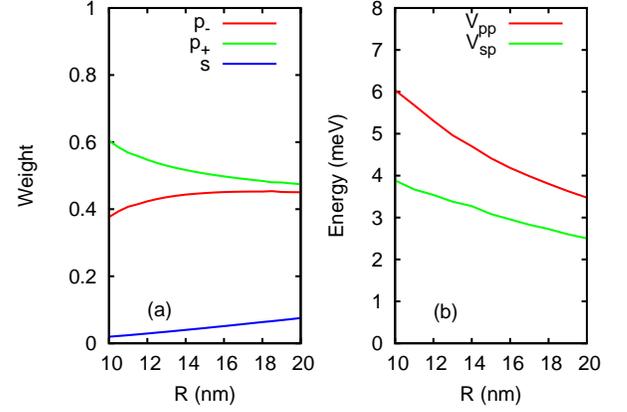}
\caption{(Color online) (a) The weights of $p$-shell ($p_{\pm}$) and $s$-shell non-interacting excitons in the fully interacting exciton
$X_{p\downarrow}$ (see Eq. (\ref{XP})) as a function of radius $R$ for CdTe QDs. (b) Non-conserving matrix elements
of the e-h interaction as a function of $R$. $V_{sp}$ mixes $s$ and $p$ shell non-interacting excitons while $V_{pp}$ mixes
the $p$-type excitons. $W=5$nm.}
\label{fig2}
\end{figure}

We find that for disk-shaped QDs the lowest energy fully interacting $p$-type exciton $X_{p\downarrow}$ is generally a mixture
of non-interacting electron-hole states:
\begin{equation}\label{XP}
|X_{p\downarrow}\rangle  =C_{\downarrow_{p_-}}|\downarrow_{p_-}\Uparrow_{p_+} \rangle
+C_{\downarrow_{p_+}}|\downarrow_{p_+}\Uparrow_{p_-}  \rangle+C_{s\downarrow}|\downarrow_s\Uparrow_s \rangle,
\end{equation}
where the $p$-shell configurations have almost equal weight while the pure
$s$-shell state has a very small contribution, i.e. $|C_{\downarrow_{p_-}}|^2\approx |C_{\downarrow_{p_+}}|^2\gg |C_{s\downarrow}|^2$.
In other words the dominant character of the state $X_{p\downarrow}$ is $p$-like. Similarly, the purity of the
interacting $s$-shell exciton is typically around $95\%$. Note that by convention the spin index of the exciton state and of the
weight coefficients refer to the electron spin orientation. We shall see in the next subsection that the Coulomb mixing also operates on biexciton states.

In the absence of electron-hole interaction (i.e for $W^{eh}=0$) we find that the gap between the states
$|\downarrow_{p_-}\Uparrow_{p_+} \rangle$ and $|\downarrow_{p_+}\Uparrow_{p_-}\rangle$ is very close to the gap $\Delta_{pp}$ within the $p$ shell,
which was found to be less than 1meV for a wide range of radii (see Fig.\,1(b)).

Fig.\,2(a) displays the dependence of the weights $|C_{\downarrow_{p_{\pm}}}|^2$ and $|C_{s\downarrow}|^2$ on QD radius.
The weight of the noninteracting $s$-shell state is very small at $R=10$nm and increases slowly to $8\%$ at $R=20$nm.
On the contrary, both $p$-shell states $|\downarrow_{p_-}\Uparrow_{p_+} \rangle$ and
$|\downarrow_{p_+}\Uparrow_{p_-}\rangle$ contribute substantially to the exciton state.
In Fig.\,2(b) we show the dependence on $R$ of the non-conserving matrix element of the interband interaction
$V_{pp}=\langle\downarrow_{p_+}\Uparrow_{p_-}|W^{eh}|\downarrow_{p_-}\Uparrow_{p_+}\rangle$ which leads to the mixing of
the two $p$-shell states. We see that the matrix element $V_{pp}$
 exceeds the gap within the $p$ shell and therefore a strong mixing is expected.
 At small $R$ the mixing of $s$ and $p$-shell states is
 weak because the gap $\Delta_{sp }$ between them (see Fig.\,1(b)) exceeds
by far the off-diagonal matrix element $V_{sp}$ which is also shown in Fig.\,2(b). As $R$ increases the ratio
$\Delta_{sp}/V_{sp}$ decreases and the $sp$ mixing is enhanced.

\begin{figure}[tbhp!]
\includegraphics[angle=0,width=0.45\textwidth]{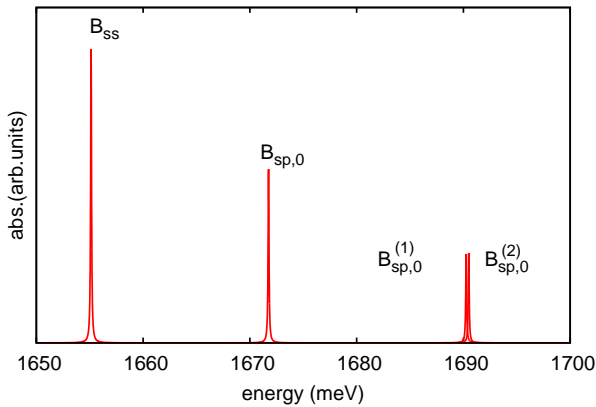}
\caption{(Color online) The absorption spectrum associated to transitions between the $s$-shell exciton $X_{s\downarrow}$
and biexciton states under a $\sigma_-$ pulse. The states containing one singlet and one triplet are clearly separated from
the $sp$ antiparallel biexciton $B_{sp,0}$ (see the text for notation).
Other parameters: $R=15$nm, $W=5$nm, $F=35$kV/cm, $\epsilon_r=7.1$ for CdTe.}
\label{fig3}
\end{figure}

Let us discuss now the structure of $sp$ biexcitons. They are conveniently described in terms of triplet and
singlet states for electrons and holes.
The $sp$ two-hole states are labelled by the total hole spin $S_z^h=0,\pm 1$:
\begin{eqnarray}\label{T-para}
|T_{sp_{\pm}}^{h,1}\rangle &=&|\Uparrow_s\Uparrow_{p_{\pm}}\rangle,\quad |T_{sp_{\pm}}^{h,-1}\rangle=|\Downarrow_s\Downarrow_{p_{\pm}}\rangle,\\
\label{T-anti}
|T_{sp_{\pm}}^{h,0}\rangle &=&\frac{1}{\sqrt 2}\left (|\Uparrow_s\Downarrow_{p_{\pm}}\rangle + |\Downarrow_s\Uparrow_{p_{\pm}}\rangle \right ),\\\label{S-anti}
|S_{sp_{\pm}}^{h,0}\rangle &=&\frac{1}{\sqrt 2}\left (|\Uparrow_s\Downarrow_{p_{\pm}}\rangle - |\Downarrow_s\Uparrow_{p_{\pm}}\rangle \right ).
\end{eqnarray}
A similar notation holds for two-electron states in terms of the total electron spin $S_z^e$ (e.g. $T_{sp_{\pm}}^{e,S_z^e}$).
The $sp$ biexciton states will be denoted by $B_{sp,S_z^e}$ and given as linear combinations of two-electron and
two-hole states introduced above. For further purpose we label the biexciton states w.r.t the total electronic spin.
The lowest energy optically active $sp$ biexcitons are given below, $\approx$ meaning that we neglected other configurations
whose weights are very small (typically less than $1\%$):
\begin{eqnarray}\label{Bsp_0}
|B_{sp,0}\rangle&\approx& \frac {1}{\sqrt 2}\left (|T^{e,0}_{sp_-}\rangle |T^{h,0}_{sp_+}\rangle+|T^{e,0}_{sp_+}\rangle |T^{h,0}_{sp_-}\rangle\right ),\\\label{Bsp_1}
|B_{sp,\pm 1}\rangle&\approx& \frac {1}{\sqrt 2} \left (|T^{e,\pm 1}_{sp_-}\rangle |T^{h,\mp 1}_{sp_+}\rangle
+|T^{e,\pm 1}_{sp_+}\rangle |T^{h,\mp 1}_{sp_-}\rangle\right).
\end{eqnarray}
Again one observes that the small gap within the $p$ shell leads to a rather complicated biexciton structure
containing triplet states from $p_{\pm}$ subshells.
The $s$-shell biexciton is simply $B_{ss}=|\uparrow_s\downarrow_s\Downarrow_s\Uparrow_s\rangle$.

Let us comment on the effect of electron-hole (e-h) exchange interaction which is neglected in the present work.
Pseudopotential calculations \cite{Bester} revealed that even for cylindrical systems
the reduced symmetry of the atomistic confining potential leads to non-vanishing e-h exchange which in turn induces
exciton fine structure. The short-range exchange is proportional to the probability for the electron and the hole
to be at the same site \cite{Romestain} and does not split the bright states. The non-local (anisotropic) part of
the e-h exchange then splits the bright states and mixes excitons with completely
opposite projections of total angular momentum (see e.g Ref. \onlinecite{Kadantsev}). This splitting is however much smaller
than the matrix element $V_{pp}$ which mixes the bright excitons from different $p$-subshells (see Fig. 2(b)).
Therefore we do not expect the weights in Eq.(\ref{XP}) to be significantly altered in the presence of the e-h exchange.
Moreover, the circularly polarized pulses used in our simulations activate excitons with well defined
spins of the electron-hole pair. Similarly, the electron-electron and hole-hole exchange interaction have even larger
values (around 18 meVs for the dots considered in our simulations) so that the $sp$-biexciton structure given
in Eqs.(\ref{Bsp_0}) and (\ref{Bsp_1}) is also stable against e-h exchange.

\subsection{All-optical manipulation of $sp$ biexcitons}

We shall now study the dynamics of $sp$ biexciton with antiparallel spins $B_{sp,0}$.
One can prepare this biexciton state by applying two laser pulses according to the following scheme:
i) a $\sigma_+$ pulse of frequency $\omega_s$ drives a $\pi$ rotation of the $s$-type exciton $X_{s\downarrow}$;
ii) the pulse is switched off once the {\it initialization} of $X_{s\downarrow}$ is complete (at instant $t=t_s$) and a 2nd $\sigma_-$
{\it control} pulse of frequency $\omega_p$ is simultaneously turned on; iii) the latter optically activates the $p$-shell hole
 states leading thus to the appearance of the $sp$ biexciton.
Note that the biexcitons with parallel spins ($S_z^e=\pm 1$) can only be prepared if the initialization and control pulses have the same
polarization, while the preparation of antiparallel configuration ($S_z^e=0$) requires opposite polarizations. This means in particular
that the degenerated biexciton states $B_{sp,0}$ and $B_{sp,\pm 1}$ can be individualy selected by appropriate polarization of the pulses.
The intraband relaxation is expected to strongly damage antiparallel spin configurations (see Eqs.(\ref{T-anti}),(\ref{S-anti}))
while the parallel triplets $T^{e,\pm 1}_{sp_{\mp}} $ are more robust due to the spin blockade effect
\cite{Kavousanaki}.

The biexciton absorption spectrum corresponding to the initial state $X_{s\downarrow}$ is shown in Fig.\,3. The 2nd $\sigma_-$
pulse scans the $s$ shell biexciton  $B_{ss}$ and various $sp$ biexcitons.
Besides $B_{sp,0}$ we find two higher energy biexcitons containing both singlet and triplet states, namely: $|B_{sp,0}^{(1)}\rangle\approx\frac {1}{\sqrt 2}(|T^{e,0}_{sp_-}\rangle |S^{h,0}_{sp_+}\rangle+|T^{e,0}_{sp_+}\rangle |S^{h,0}_{sp_-}\rangle)$ and $|B_{sp,0}^{(2)}\rangle\approx\frac {1}{\sqrt 2}(|S^{e,0}_{sp_-}\rangle |T^{h,0}_{sp_+}\rangle+|S^{e,0}_{sp_+}\rangle |T^{h,0}_{sp_-}\rangle)$.
The {\it electron} singlet-triplet splitting $2J_{{\rm ee}}$ can be extracted from Fig.\,3  as the energy difference
${\cal E}(B_{sp,0}^{(2)})-{\cal E}(B_{sp,0})$.
We find $2J_{ee}$ to be around 18meV, in very good agreement with the data reported in Ref.\onlinecite{Kazimierczuk1}.
Note that the ratio between the amplitudes of the absorption peaks for singlet
($B_{sp,0}^{(2)})$ and triplet ($B_{sp,0}$) biexcitons is around 0.5, as observed in experiment \cite{Kazimierczuk1}.

\begin{figure}[tbhp!]
\includegraphics[angle=0,width=0.45\textwidth]{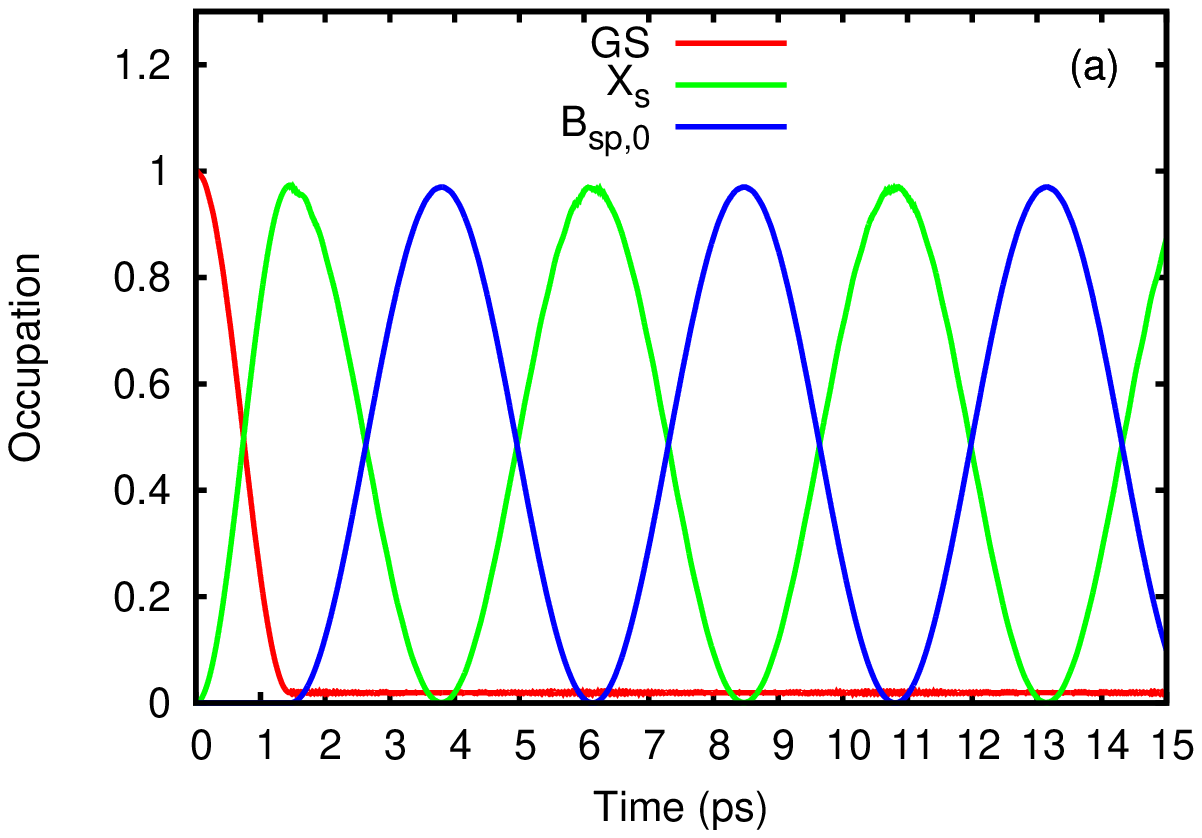}
\includegraphics[angle=0,width=0.45\textwidth]{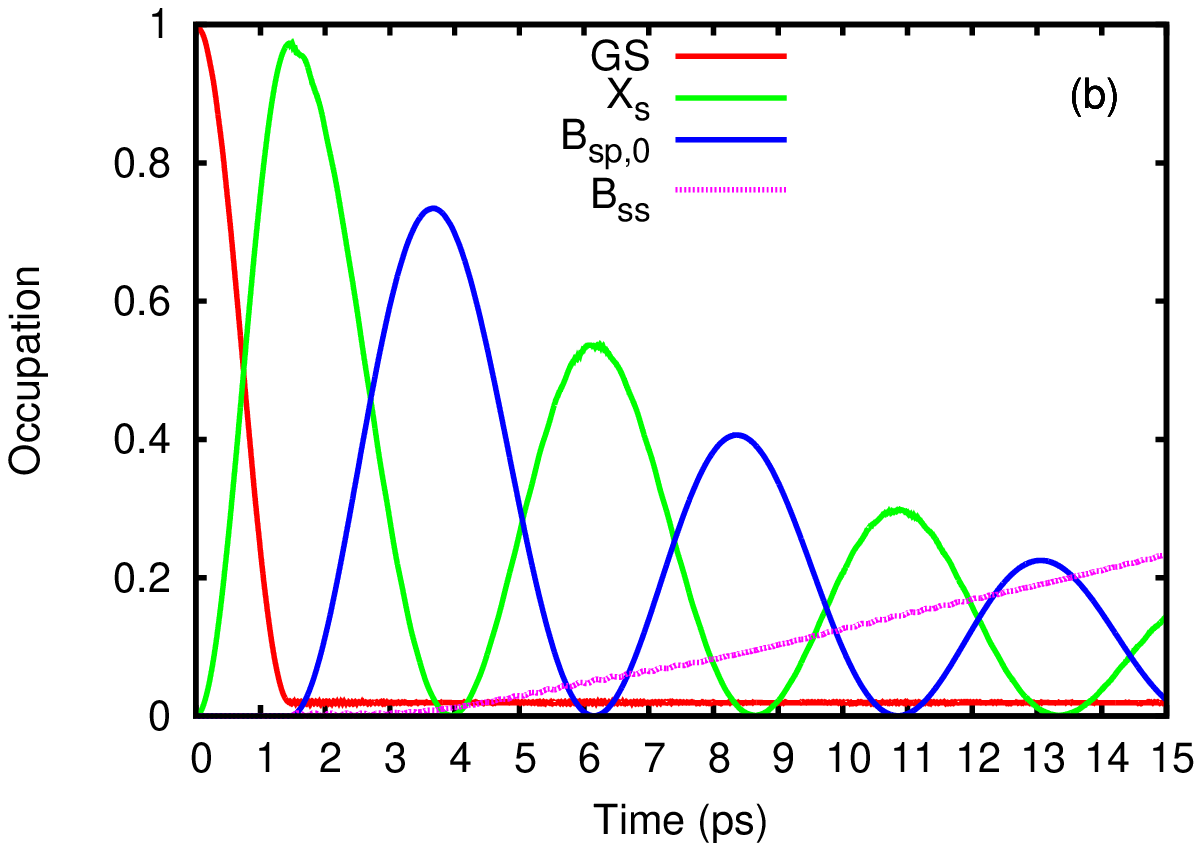}
\caption{(Color online)
The Rabi oscillations of the excitonic and biexcitonic populations $P(X_{s\downarrow})$
and $P(B_{sp,0})$ with and without intraband relaxation.
The 1st $\sigma_+$ pulse initializes the control exciton state $X_{s\downarrow}$
and is turned off at $t_s=1.45$ps. The 2nd $\sigma_-$ pulse starts at the same time.
(a) C-NOT gate in the absence of IBR. (b) Relaxation effects for $\tau_e=30$ps, $\tau_h=4$ps.
The occupations of the ground state (GS) and $s$-shell biexciton $B_{ss}$ are also shown. Other parameters:
$R=15$nm, $W=5$nm, $F=35$kV/cm, $\hbar\omega_s=1.661$eV, $\hbar\omega_p=1.672$eV.}
\label{fig4}
\end{figure}

In Fig.\,4 we present the conditional dynamics of the antiparallel $sp$ biexciton generated in a QD of radius $R=15$nm
and height $W=5$nm by the two-pulse sequence described above.
Fig.\,4(a) confirms that in the absence of intraband relaxation the exciton and biexciton populations
exhibit Rabi oscillations, the QD state being periodically switched between $X_{s\downarrow}$ and $B_{sp,0}$. We also note that
the occupation sum $P(X_{s\downarrow})+P(B_{sp,0})\approx 1$ at all times so that the dynamics of the system is entirely described
 by these two states only.

\begin{figure}[tbhp!]
\includegraphics[angle=-0,width=0.45\textwidth]{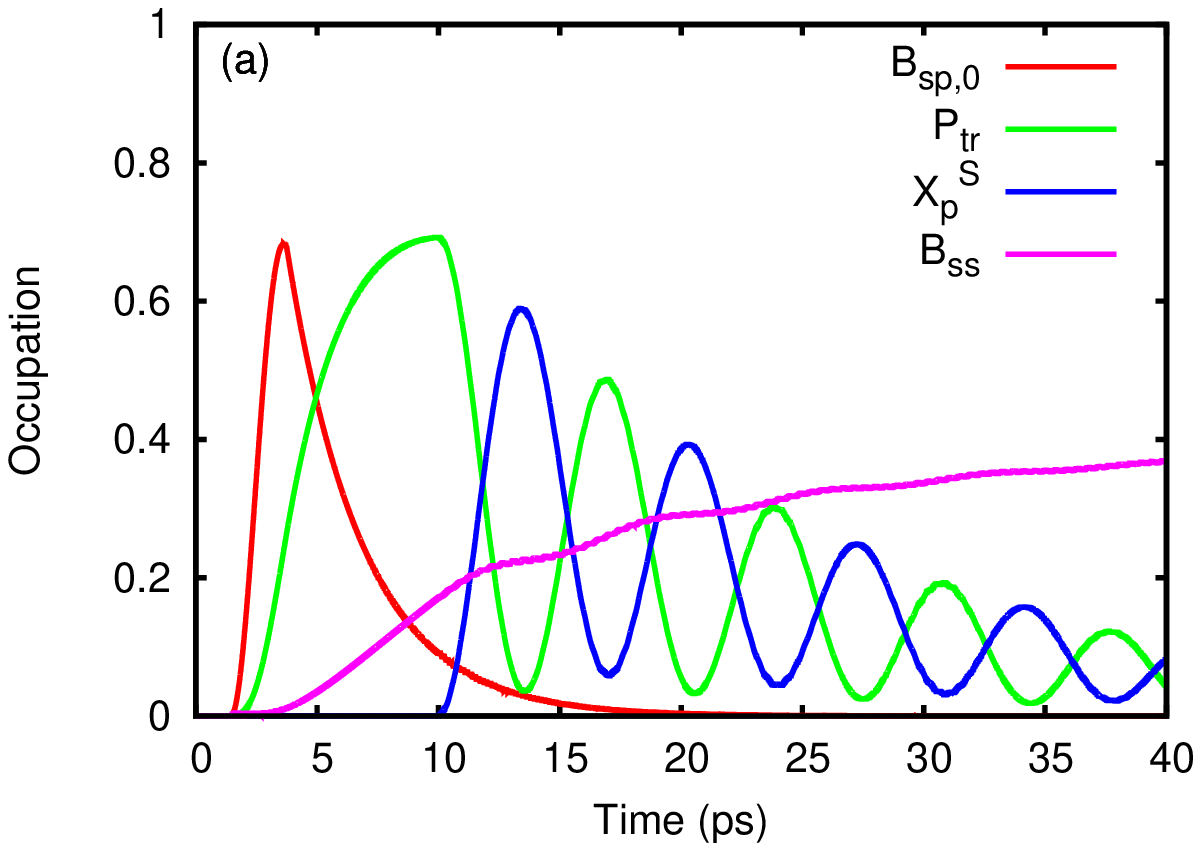}
\includegraphics[angle=-0,width=0.45\textwidth]{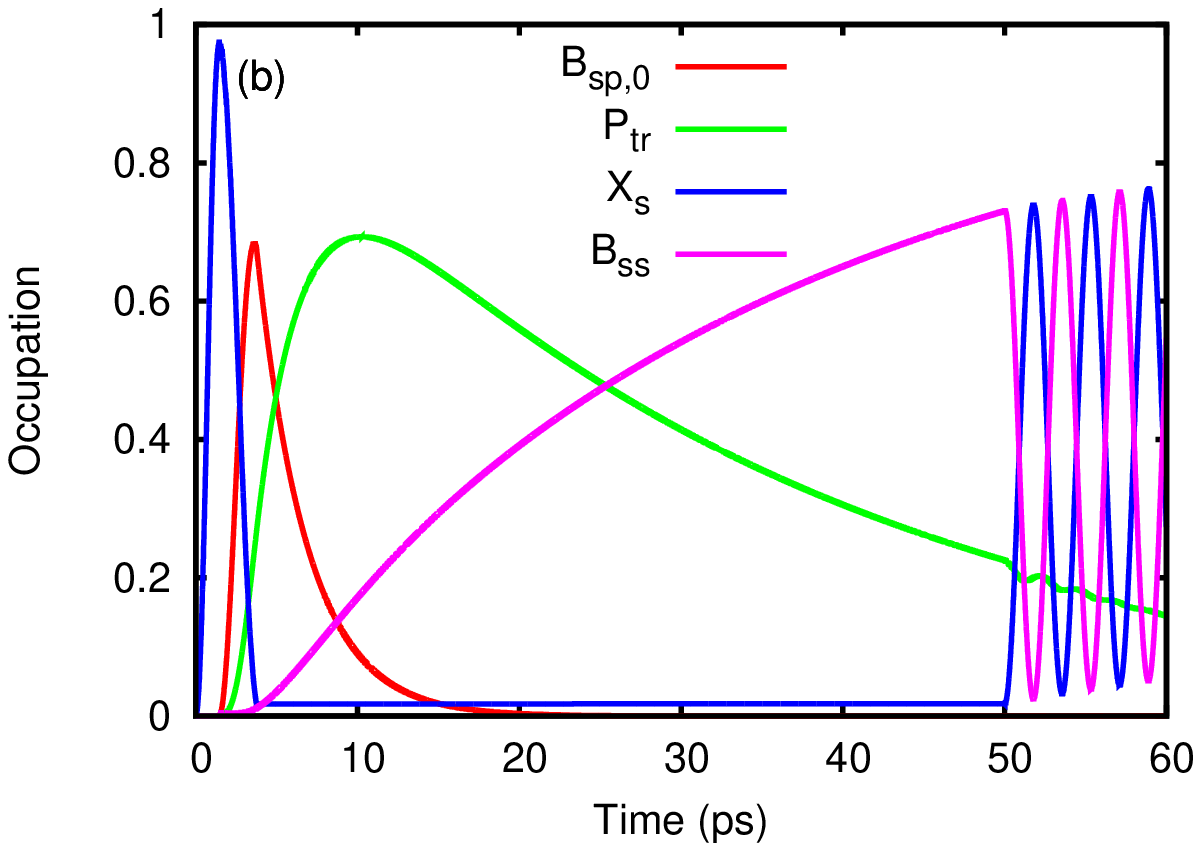}
\caption{(Color online) Population dynamics under three-pulse operation described in the text.
(a) The 3rd $\sigma_-$ pulse applied at $t=10$ps generates damped Rabi oscillations of the total occupation of the
transient states (see the definition of $P_{{\rm tr}}$ in Eq.(\ref{TS})) and of the skewed exciton states.
The population of the $sp$ biexciton vanishes around $t=17$ps.
(b) Rabi oscillations of the $s$ shell-exciton and biexciton due to a 3rd pulse of frequency $\omega_B$ applied at $t=50$ps.
Other parameters: $R=15$nm, $W=5$nm, $\tau_e=30$ps, $\tau_h=3$ps, $F=35$kV/cm,
$\hbar\omega_B=1.654$eV, $\hbar\omega_X=1.645$eV. }
\label{fig5}
\end{figure}

The two-color protocol described above resembles the conditional C-NOT gate proposed in Ref.\onlinecite{Troiani},
the difference being that here we deal with $sp$ biexcitons instead of $s$-shell biexcitons.
More precisely, the control state is the $s$-shell exciton while the target state is the $sp$ biexciton generated by the 2nd control pulse.
The conditional feature comes from the fact that the biexciton population changes noticeably if and only if the $s$-shell
exciton is occupied. This imposes thus a resonant condition for the 2nd pulse $\hbar\omega_p\sim {\cal E}(B_{sp,0})-{\cal E}(X_{s\downarrow})$.
Note that the 'target' biexciton state cannot be written as a product of two excitonic qubits (see Eq.(\ref{Bsp_0})); this would be the case
if one restricts the dynamics to $s$-shell excitons. As a consequence the relevant computational space \cite{Biolatti2} for the C-NOT operations
is more complicated.
 For all simulations presented below we have checked that the formation of more complicated MBS (i.e three e-h pairs)
due to simultaneous transition of two electrons from the VB $p$-shell to CB can be safely neglected.
The dynamics of paralel triplet states $B_{sp,\pm 1}$ is rather similar to the one displayed in Fig.\,4(a), because the
spin Coulomb blockade prevents fast relaxation processes (not shown).

The effect of intraband relaxation processes
is shown in Fig.\,4(b). The Rabi oscillations are strongly damped and the 2nd maximum of the biexciton population does not exceed $45\%$,
therefore the fidelity of the exciton-biexciton C-NOT gate is rapidly compromised.
One also notices the uniform filling of the $s$-type biexciton  $B_{ss}$
due to the electron intraband relaxation. In the long time limit $P(B_{ss})$ saturates at $85\%$ (not shown).

We now discuss in more detail the relaxation processes. Already by looking at Fig.\,4(b) one can guess that in this case the
dynamics does not involve only the states $X_{s\downarrow}$, $B_{sp,0}$ and $B_{ss}$; if this was the case one
would check that the occupations of these states sum up to 1.
It is therefore clear that the QD relaxation to the biexciton state $B_{ss}$ activates some transient states (TS).
One can see for example that the $s\to p$ spin-conserving relaxation drives the system to a final $s$-shell biexciton through
the following 'paths':
\begin{eqnarray}\label{fast}
&&|B_{sp,0}\rangle\Longrightarrow |T^{e,0}_{sp_{\pm}}\rangle |\Uparrow_s\Downarrow_s\rangle\longrightarrow |B_{ss}\rangle,\\\label{slow}
&&|B_{sp,0}\rangle\longrightarrow |\uparrow_s\downarrow_s\rangle |T^{h,0}_{sp_{\pm}}\rangle \Longrightarrow |B_{ss}\rangle .
\end{eqnarray}
Here the double (simple) arrow marks relaxation processes of holes (electrons).
Since the hole relaxation is the fastest process the paths given in Eq.(\ref{fast}) are more favorable
and dominate the dynamics on short time scale (this fact wil be confirmed by numerical results presented sligthly below).
Clearly the antiparallel triplet configurations $|T^{e,0}_{sp_{\pm}}\rangle|\Downarrow_s\Uparrow_s\rangle$ become populated via
 hole triplet relaxation whereas $|\uparrow_s\downarrow_s\rangle |T^{h,0}_{sp_{\pm}}\rangle$ are filled
via slower electron triplet relaxation. Note also that the hole relaxation within the $p$ shell
(i.e. $\Uparrow_{p_+}\to\,\Uparrow_{p_-}$ and $\Downarrow_{p_-}\to\,\Downarrow_{p_+}$ ) leads to
the occupation of the states $|T^{e,0}_{sp_{\pm}}\rangle|T^{h,0}_{sp_{\pm}}\rangle $ which will further
relax to $|T^{e,0}_{sp_{\pm}}\rangle |\Uparrow_s\Downarrow_s\rangle$.
It is useful to define $P_{{\rm tr}}$ as the sum of the occupations of the four transient states introduced above
(see the middle states in Eqs.(\ref{fast}) and (\ref{slow})):
\begin{eqnarray}\nonumber
P_{{\rm tr}}&=&P(|T^{e,0}_{sp_{+}}\rangle |\Uparrow_s\Downarrow_s\rangle)+P(|T^{e,0}_{sp_{-}}\rangle |\Uparrow_s\Downarrow_s\rangle )\\\label{TS}
&+&P(|\uparrow_s\downarrow_s\rangle |T^{h,0}_{sp_{+}}\rangle)+P(|\uparrow_s\downarrow_s\rangle |T^{h,0}_{sp_{-}}\rangle).
\end{eqnarray}
The transient states $|T^{e,0}_{sp_{\pm}}\rangle|\Downarrow_s\Uparrow_s\rangle$
are non-trivial due to the exchange-coupled electrons belonging to different shells. Moreover, these states can be probed
with pulses centered on the $s$-shell transition, which allows us to get information on their lifetime and occupation.
We find that under a 3rd $\sigma_-$ pulse the states $|T^{e,0}_{sp_{\pm}}\rangle|\Downarrow_s\Uparrow_s\rangle$ couple to $p$-like skewed excitons
$|X_{p_{\pm}}^{S}\rangle =1/\sqrt {5}|\downarrow_s\Uparrow_{p_{\pm}}\rangle+2/\sqrt {5}|\downarrow_{p_{\pm}}\Uparrow_s\rangle$
(the electronic $p$ character of these states comes from the dominant weight of $|\downarrow_{p_{\pm}}\Uparrow_s\rangle$).

Based on the above observations we simulate in Fig.\,5 a three-pulse experiment for the optical probing
of the transient states. The setup goes as follows:
i) The antiparallel $sp$ biexciton is generated as before but now the 2nd pulse is turned off once the population $P(B_{sp,0})$
reached its maximum (i.e. at $t=3.75 $ps). ii) The system relaxes through TS without
any optical driving. iii) At $t=10$ps we tune a 3rd $\sigma_-$ pulse to the transitions between TS and skewed excitons $X_{p_{\pm}}^{S}$.
The pulse frequency is set as
$\hbar\omega_X={\cal E}(|T^{e,0}_{sp_{\pm}}\rangle|\Downarrow_s\Uparrow_s\rangle)-{\cal E}(X_{p_{\pm}}^{S})$.
The dynamics of the system along this procedure is not difficult to read from Fig.\,5.
First, the $sp$ biexciton displays a clear exponential decay.
From Fig.\,5(a) one notices that at $t=10$ps $P_{{\rm tr}}$ is around $69\%$
while the occupation of $B_{ss}$ is only around $18\%$. It is therefore clear that the TS are filled much faster than the $s$-shell biexciton,
the latter being populated via slower electron relaxation. The optical response of the transient states is recorded as damped Rabi oscillations
for $t>10$ps. We also plot the total occupation of skewed states. The Rabi oscillations damping comes obviously from the relaxation of the $p$-shell electrons.

One should realize that in order to get a strong optical response of the transient states a certain delay $\Delta t$ between the
2nd and the 3rd pulses is needed in order to achieve a significant filling. Typically one should consider
$\Delta t>\tau_h$ (in Fig.\,5(a) $\Delta t=6.25$ps). On the other hand, if $\Delta t$
is sufficiently large the observed Rabi oscillations will rather originate from the transition between the biexciton $B_{ss}$
and the $s$-shell excitons. The resonant frequency $\hbar\omega_B={\cal E}(B_{ss})-{\cal E}(X_{s\downarrow})$.
We illustrate this fact in Fig.\,5(b) where the 3rd pulse is applied at $t=50$ps ($\Delta t=46.25$ps)). At that instant the occupation of the
TS considerably diminished and the $s$-type biexciton is the dominant configuration.
The periodic oscillations cannot be associated to TS as their occupation continues to decrease slowly and do not respond to the pulse.

On the other hand clear oscillations of the $s$-shell exciton develop. Let us stress here that the frequencies
$\omega_{B}$ and $\omega_{X}$ are actually very close (see the values given in the caption of Fig.\,5 for the present simulation).
However, one can still identify the origin of the Rabi oscillations from their aspect, i.e. damped (when associated to transient states)
or not (when given by pure $s$-shell configurations).

\begin{figure}[tbhp!]
\includegraphics[width=0.525\textwidth]{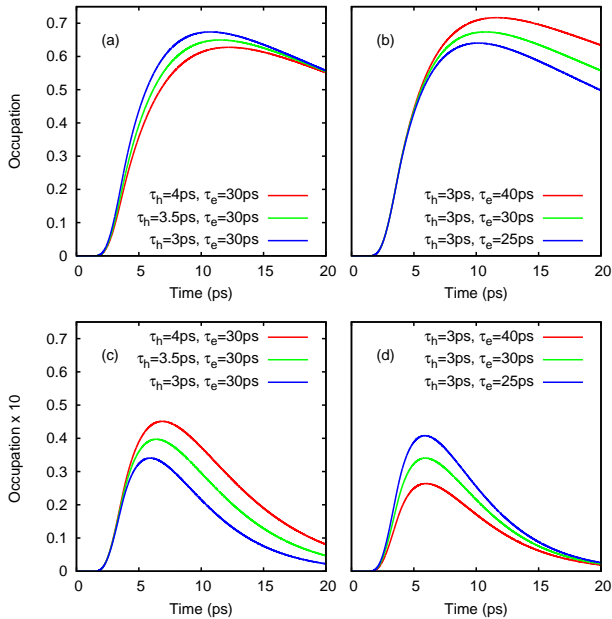}
\vspace{4cm}
\caption{(Color online) (a) The total occupation $P_{{\rm fast}}$ of transient states containing antiparallel
$sp$ electronic triplets for different hole relaxation times $\tau_h$. (b) The same quantity for different electron relaxation times $\tau_e$.
 The slow total filling $P_{{\rm slow}}$ as function of $\tau_h$ (c) and $\tau_e$ (d).
Note that in Figs.\,(c) and (d) the occupations are multiplied by 10. $P_{{\rm fast}}$ and $P_{{\rm slow}}$ are defined in the text.
Other parameteres: $R=15$nm, $W=5$nm, $\tau_h=3$ps, $F=35$kV/cm. }
\label{fig6}
\end{figure}

The dynamics of the transient states was further investigated by performing numerical simulations for
different relaxation times $\tau_e$ and $\tau_h$. We plot in Figs.\,6(a) and (b) the total occupation of the antiparallel
triplet transient states $P_{{\rm fast}}=P(|T^{e,0}_{sp_+}\rangle|\Downarrow_s\Uparrow_s\rangle)+P(|T^{e,0}_{sp_-}\rangle|\Downarrow_s\Uparrow_s\rangle)$ which are filled through fast hole relaxation processes.
For completeness we also present in Figs.\,6(c) and (d) the occupation
$P_{{\rm slow}}=P(|\uparrow_s\downarrow_s\rangle |T^{h,0}_{sp_+}\rangle)+P(|\uparrow_s\downarrow_s\rangle |T^{h,0}_{sp_-}\rangle)$
of the slowly filled states.
The setup is the same as in Fig.\,5, namely the 2nd pulse is turned off
at $t=3.75$ps. The occupation $P_{{\rm fast}}$ exceeds by far the filling of the states $|\uparrow_s\downarrow_s\rangle |T^{h,0}_{sp_{\pm}}\rangle$ which requires spin relaxation.
The maximum of $P_{{\rm fast}}$ increases when  $\tau_h$ decreases or if $\tau_e$ increases (slower electron relaxation).
The three curves shown in Fig.\,6(a) overlap for $t>20$ps as the spin relaxation time is the same.
In contrast, the short time dynamics does not change with $\tau_e$ (see  Fig.\,6(b)), confirming that electron relaxation processes
are not important in this regime.
The 'slow' states behave quite differently, namely their maximum increases if  $\tau_h$ increases or
when $\tau_e$ decreases.
 Note that $P_{{\rm slow}}$ depends on $\tau_h$ at short times because of the triplet hole
states $T^{h,0}_{sp_{\pm}}$.
It is also clear that the fast transient states dominate the dynamics, the contribution of the 'slow'
ones being negligible and vanishing at longer times.

The above results confirm that in the presence of hole relaxation one cannot neither prepare a long-lived $sp$
biexciton state nor achieve a high-efficiency C-NOT gate. We will show instead that the antiparallel $sp$ triplet
configuration of the transient states is more stable against intraband relaxation and can be therefore used as two-qubit
electron state in further manipulation schemes.

 Note that $P_{{\rm slow}}$ depends on $\tau_h$ at short times because of the triplet hole
states $T^{h,0}_{sp_{\pm}}$.
It is also clear that the fast transient states dominate the dynamics, the contribution of the 'slow'
ones being negligible and vanishing at longer times.

\begin{figure}[tbhp!]
\includegraphics[width=0.45\textwidth]{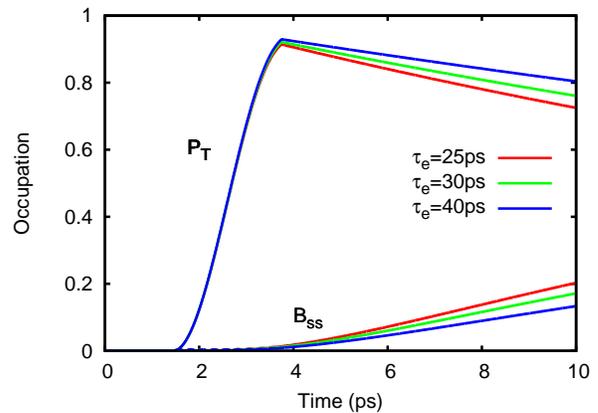}
\caption{(Color online)  The total occupation $P_{T}$ of antiparallel electronic $sp$ triplet at different relaxation
times $\tau_e$. The occupation of the biexciton state $B_{ss}$ increases faster
when the electron relaxation time decreases. Other parameters: $R=15$nm, $W=5$nm, $\tau_h=3$ps, $F=35$kV/cm. }
\label{fig7}
\end{figure}

The above results confirm that in the presence of hole relaxation one cannot neither prepare a long-lived $sp$
biexciton state nor achieve a high-efficiency C-NOT gate. We will show instead that the antiparallel $sp$ triplet
configuration of the transient states is more stable against intraband relaxation and can be therefore used as two-qubit
electron state in further manipulation schemes.

The main point is that the $sp$ biexciton and the transient states differ only by the configuration of holes,
whereas the electrons occupy one of the triplet states $T^{e,0}_{sp_{\pm}}$. As a consequence, the time-dependent occupation
of antiparallel triplet configuration is given by $P_{T}=P(B_{sp,0})+P_{{\rm fast}}$.
We plot this quantity in Fig.\,7 for different electron relaxation times. One observes that the occupation of
the antiparallel triplet configurations in the conduction band is not significantly affected as long as the filling of the $s$-shell
biexciton via electron relaxation processes is sufficiently small (say less than $10\%$).
The latter is also shown in Fig.\,7 for the same parameters. Looking at Fig.\,7 we see that the probability
of antiparallel configuration of the conduction band exceeds $85\%$ for few picoseconds at $\tau_e=25$ps, while for
$\tau_e=40$ps  $P_{T}$ is still large (i.e $80\%$) even at $t=10$ps or $t=15$ps ($70\%$).

Otherwise stated, the proposed two-pulse sequence can be used to prepare antiparallel $sp$ triplet configurations for electrons,
without taking care of the hole configuration. Let us mention here that the triplet state measured in Ref.(\onlinecite{Kazimierczuk1})
appears only if the initial QD state is an excited one: there are two extra electrons in the conduction band and no holes (full valence band).
In our setup the initial state is the ground state (empty conduction band and full valence band) which allows a better control over the
fast initialization of the triplet state.
We stress that in the present two-color protocol the antiparallel triplet states appear due to electron-electron exchange interaction.
According to the single-particle optical selection rules the 2nd $\sigma_-$ pulse would only generate the electron-hole pair
$(\uparrow_{p_{\pm}},\Downarrow_{p_{\mp}})$. This is actually the case if the $\sigma_-$ pulse acts on the ground state.
However, in our scenario the 2nd pulse finds the system in the exciton state $X_{\downarrow_s}$ and the
exchange interaction instantaneously leads to the formation of antiparallel triplet state.
We have performed numerical simulations for QDs of different sizes and obtained similar results.

Finally we emphasize that the role of electron-electron exchange interaction in CdTe QDs cannot be neglected, its value
(see e.g Ref.\onlinecite{Kazimierczuk1}) being much larger than for InAs/GaAs QDs (few meV). In the latter case
one can argue \cite{Kavousanaki} that if the pulse width exceeds the singlet-triplet splitting the exchange interaction effects
are washed out. The setup studied here involves rather sharp pulses, the goal of the two-pulse protocol being to prepare a given
non-trivial many-body state rather than probing the relaxation processes.

\section{Conclusions}

We investigated fast conditional operations on $sp$ biexcitons in CdTe disk-shaped QDs
in the presence of intraband carrier relaxation.
The single-particle spectral properties are obtained from the four-band $k\cdot p$ theory,
whereas the Coulomb interaction is taken into account within the configuration-interaction method.
 The von Neumann-Lindblad equation is used to derive the time-dependent occupation of relevant exciton and biexciton states.
The disk-shaped QDs have a rich spectral single-particle structure which leaves its fingerprints on the exciton and
biexciton states. In particular we find that the small gap within the $p$ shell leads to strong configuration
mixing for both exciton and biexciton states.

We focused on two-pulse manipulation schemes activating first an $s$-shell exciton and subsequently an $sp$ biexciton.
As expected, the exciton-biexciton Rabi oscillations are strongly damped if the target biexciton state is made of triplet
states with antiparallel spins. At short times the main source of decoherence comes from fast hole relaxation processes.
More precisely, the relaxation from the $sp$ state is mediated by transient states which then slowly
deplete via electron relaxation in favor of the final $s$-shell biexciton state. The transient states are
optically active and also lead to damped optical Rabi oscillations under a 3rd probe pulse.

Our simulations show that the electronic antiparallel triplet configurations present in the transient states
preserve a rather high purity for at least 5ps after the pump pulse is switched off.
This suggests that appropriate two-color manipulation protocols can be used to initialize antiparallel $sp$
triplet states in the conduction band. Such states are no-trivial and may provide information on the dynamics of entangled electrons.
It also turns out that in the presence of intraband relaxation the dynamics of the $sp$ biexcitons cannot be
described by an effective $4\times 4$ evolution operator as it is usually the case in the study of pure
$s$ type biexcitons \cite{Troiani} or for two-qubit operations in Coulomb-coupled quantum dots \cite{Danckwerts}.
This happens because the intraband relaxation activates configurations which are otherwise forbidden, increasing thus the
Fock subspace relevant for the QD dynamics.

\begin{acknowledgments}
The authors acknowledge financial support from PNCDI2 program (grant PN-II-ID-PCE-2011-3-0091) and from grant No.\ 45N/2009.
We thank Paul Gartner and Paul Racec for illuminating discussions.
\end{acknowledgments}

\end{document}